\documentstyle[amstex,a4,12pt,epsfig,here]{article}
\begin{document}
\title{A Search for Vector Diquarks at the CERN LHC}
\author{E. Arik$^{a}$, O. \c{C}ak\i r$^{b}$, S. A. \c{C}etin$^{a}$ and S.
Sultansoy$^{c,d}$ \\
$^a$ Bogazici University, Faculty of Sciences,
Department of Physics,\\
80815, Bebek, Istanbul, Turkey\\
\,$^b$ Ankara University, Faculty of Sciences,
Department of Physics, \\
06100 Tandogan, Ankara, Turkey\\
\,$^c$ Gazi University, Faculty of Arts and Sciences,
Department of Physics,\\
06500, Teknikokullar, Ankara, Turkey\\
\,$^d$ Azerbaijan Academy of Sciences, Institute of Physics,\\
H. Cavid av., 33, Baku, Azerbaijan}

\date{ }

\maketitle

\begin{abstract}
Resonant production of the first generation vector diquarks 
 at the
CERN Large Hadron Collider (LHC) is investigated. It is
shown that the LHC will be able to discover vector diquarks with
masses up to 9 TeV for quark-diquark-quark coupling $\alpha_D=0.1$ and 
4 TeV for  $\alpha_D=5 \times 10^{-4}$.
\end{abstract}

\vskip 0.5cm

The existence of at least three fermion families and especially inter-family 
mixings 
naturally lead to the hypothesis that they are made up of more fundamental 
constituents frequently called preons \cite{terazawa80}. Today, the 
compositeness should be
considered as a candidate for beyond the standard model (BSM) physics on 
the same footing as SUSY. Moreover, some
assumptions made in order to get rid of huge number of free parameters in 
three family  minimal supersymmetric standard model (MSSM) have
natural explanation (see \cite{sultansoy00} and references therein) in 
the framework of preonic models. These models
predict a rich spectrum of new particles with unusual quantum
numbers at high energies such as excited quarks and leptons,
diquarks, dileptons, leptoquarks, leptogluons, sextet quarks,
octet bosons etc. In preonic models, diquarks are
as natural as leptoquarks \cite{wudka86}.
They are colored objects having integer-spin and baryon number
$\vert B\vert= 2/3$ or $0$. Diquarks are also predicted in the
framework of superstring-inspired $E_{6}$ models \cite{hewett89}.

Recent collider limits on the diquark masses come from Tevatron
data which exclude the region 290 $< M_D < \, $420 GeV (for E$_{6}$
diquarks) \cite{abe97}. Diquark production at $e^{+}e^{-}$
colliders, $ep$ colliders and $p\bar{p}$ colliders  have been
analyzed in \cite{schaile87}, \cite{rizzo89} and \cite{angel87}
respectively. The resonance production of scalar
diquarks at CERN LHC has been studied in \cite{atag98}.

In this work, we study composite vector diquark production at the  LHC.  
Interaction Lagrangian and quantum numbers of diquarks are discussed. 
Decay width and production cross section of vector diquarks at the LHC 
are considered. Vector diquark signal and corresponding backgrounds are 
analyzed. 

A model independent, baryon number conserving, most general
$SU(3)_{C}\times SU(2)_{W}\times U(1)_{Y}$ invariant effective
Lagrangian for diquarks has the form
\begin{eqnarray}
L_{|B|=0} &=&f_{1L}\overline{q}_{L}\gamma ^{\mu }q_{L}D_{1\mu
}^{c}+(f_{1R} \overline{d}_{R}\gamma ^{\mu }d_{R}+f_{1R}^{\prime
}\overline{u}_{R}\gamma
^{\mu }u_{R})D_{1\mu }^{\prime c}  \nonumber \\
&&+\widetilde{f}_{1R}\overline{u}_{R}\gamma ^{\mu
}d_{R}\widetilde{D}_{1\mu }^{c}+f_{3L}\overline{q}_{L}{\bf {\tau}}
\gamma ^{\mu }q_{L}\cdot {\bf D}
_{3\mu }^{c} \nonumber \\
&&+f_{2}\overline{q}_{L}i\tau
_{2}u_{R}D_{2}^{c}+
\widetilde{f}_{2}\overline{q}_{L}i\tau_{2}d_{R}\widetilde{D}_{2}^{c}
+ H.c.  \\
&& \nonumber \\
L_{|B|=2/3} &=&(g_{1L}\overline{q}_{L}^{c}i\tau
_{2}q_{L}+g_{1R}\overline{u}
_{R}^{c}d_{R})D_{1}^{c}+\widetilde{g}_{1R}\overline{d}_{R}^{c}d_{R}
\widetilde{D}_{1}^{c}  \nonumber \\
&&+\widetilde{g}_{1R}^{\prime
}\overline{u}_{R}^{c}u_{R}\widetilde{D} _{1}^{\prime
c}+g_{3L}\overline{q}_{L}^{c}i\tau _{2}{\bf {\tau} }q_{L}\cdot
{\bf D}_{3}^{c}  \nonumber \\
&&+g_{2}\overline{q}_{L}^{c}\gamma ^{\mu }d_{R}D_{2\mu
}^{c}+\widetilde{g} _{2}\overline{q}_{L}^{c}\gamma ^{\mu
}u_{R}\widetilde{D}_{2\mu }^{c}+ H.c.
\end{eqnarray}
Diquarks with baryon number $B=0$ are familiar fields, as they resemble the
electroweak gauge vectors, and the neutral and charged Higgs
scalars. Here, we consider only  $|B| = 2/3$ vector diquarks. 
 In Eq. (2), $q_L=(u_L,d_L)$ and
$q^{c}=C\bar{q}^{T}$ ($\bar{q}^c=-q^TC^{-1}$). For the sake of
simplicity, color and generation indices are omitted. Scalar
diquarks $D_{1}$, $\tilde{D}_{1}$, $\tilde{D}_{1}^{\prime}$ are
$SU(2)_{W}$ singlets and ${\bf D}_{3}$ is $SU(2)_{W}$ triplet.
Vector diquarks $D_{2}$ and $\tilde{D}_{2}$ are $SU(2)_{W}$
doublets. Diquarks may transform as anti-triplet or sextet under
$SU(3)_{C}$. At this stage, we assume that each SM generation has
its own diquarks and couplings in order to avoid flavour changing
neutral currents.

Lagrangian (2) can be rewritten in the following more transparent
form:

\begin{eqnarray}
L&=& L_S+L_V\\
\nonumber
L_{S}&=& \left[ g_{1L}\left(
\bar{u}^{c}P_{L}d-\bar{d}^{c}P_{L}u\right)
+g_{1R}\bar{u}^{c}P_{R}d\right]
D_{1}+\tilde{g}_{1R}\bar{d}^{c}P_{R}d\tilde{D}_{1} \nonumber  \\
&+&\tilde{g}^{\prime }_{1R}\bar{u}^{c}P_{R}u\tilde{D}^{\prime }_{1} 
+ \sqrt{2}g_{3L}\bar{u}^{c}P_{L}uD_{3}^{+}
-\sqrt{2}g_{3L}\bar{d}^{c}P_{L}dD_{3}^{-}\nonumber  \\
&-&g_{3L}\left( \bar{u}^{c}P_{L}d+\bar{d}^{c}P_{L}u\right) D_{3}^{0}
+H.c. \\
L_{V}&=&g_{2}\bar{u}^{c}\gamma ^{\mu }P_{R}dD_{2\mu }^{1c}
+g_{2}\bar{d}^{c}\gamma ^{\mu }P_{R}dD_{2\mu }^{2c} 
+\tilde{g}_{2}\bar{u}^{c}\gamma ^{\mu }P_{R}u\tilde{D}_{2\mu
}^{1c} \nonumber \\
&+&\tilde{g}_{2}\bar{d}^{c}\gamma ^{\mu }P_{R}u\tilde{D}_{2\mu
}^{2c}+ H.c.
\end{eqnarray}
where $P_L=(1-\gamma_5)/2$ and $P_R=(1+\gamma_5)/2$.
A general classification of the first generation,
color $\bar{3}$ diquarks is shown in Table ~\ref{table1}.

\begin{figure}[htb]
  \begin{center}
  \epsfig{file=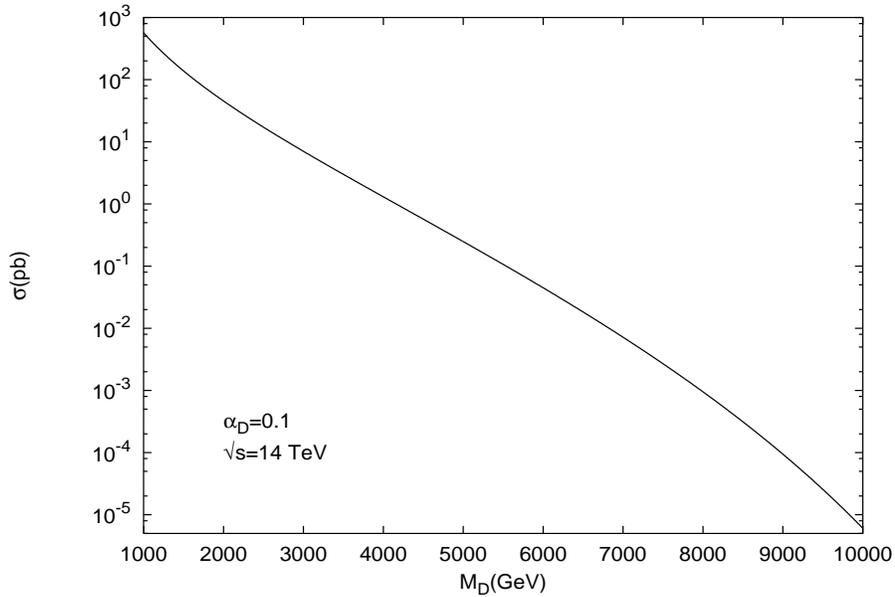,width=12cm,height=8cm}
  \end{center}
  \caption{Total cross section versus
diquark mass for $\alpha_D=0.1$.}  \label{fig1}
\end{figure}

For the present analysis, we consider the color $\bar{3}$
vector diquark $D_{2}$ with charge $1 / 3$ which couples to $ud$-pair as 
described by the effective Lagrangian (5). The decay width $\Gamma_{D}$,
derived from the same Lagrangian, is

\begin{eqnarray}
\Gamma_{D}={\alpha_D M_{D}\over 9}&\simeq &11\, GeV
({M_{D}\over {1\, TeV}})
 \;\;\; \mbox{for}\;\; \alpha_D=0.1 \qquad ,
\end{eqnarray}
where  $\alpha_D = g_{2}^{2}/4\pi$, $M_{D}$ is the mass of vector diquark.
We use the narrow width approximation and consider this as long as 
$\Gamma_D/M_D < 0.1$. The cross section
of the s-channel diquark resonance production can be
obtained as

\begin{eqnarray}
\sigma(pp\to D_{2}X)= \sigma_0 \, \int_{M_{D}^{2}/s}^{1}
{dx\over x} f_{u}(x, Q^2)f_{d}({M_{D}^{2}\over sx}, Q^2) 
\end{eqnarray}
with 
\begin{equation}
\sigma_0 = {8 \,\pi^2 \alpha_D \over 9 \, s}
\end{equation}
where the factorization scale $Q^2 = M_D^2$, $f_{u}$ and $f_{d}$ are 
quark distribution functions of each proton. 
In Fig.~\ref{fig1}, using
the CTEQ5L quark distribution functions \cite{cteq5l}, the cross
section versus diquark mass is plotted for LHC energy
($\sqrt{s}=14$ TeV) with $\alpha_D=0.1$.

$D_2$-type diquark will decay  via  $D_{2}\to u d$. 
Therefore, the signal will contain
two hard jets in the final state. 
At LHC energy, major QCD  processes contributing to two jet ($jj$)
final states and their integrated cross sections  are given in
Table \ref{table2}. The values in Table \ref{table2} have been
generated by PYTHIA 6.1 \cite{sjostrand} at parton level with
various $p_{T}$ cuts. It is clear that higher $p_T$ cuts reduce
the background cross sections significantly. These $p_T$ cuts can
be translated into the rapidity cuts via the relation between the
$p_T$ of a jet and the rapidity $y$ given by $p_T=M_{jj}/2\cosh
y$.

The differential cross section as a function of
the dijet invariant mass $M_{jj}$, with the rapidity cut
$|y_{1,2}|\le Y$, where $y_1$ and $y_2$ are rapidities of the
massless final quarks, is given by

\begin{eqnarray}
{d\sigma \over dM_{jj}}&&={M_{jj}^{3}\over 2s}
\int_{-Y}^{Y} dy_{2}
\int_{y_{1}^{\min}}^{y_{1}^{\max}} dy_{1} \;
{1\over {\cosh^{2}{y^{\star}}}} \\ \nonumber
&&\times [f_{u/A}(x_{u},Q^2)f_{d/B}(x_{d},Q^2)
{d\hat{\sigma}\over d\hat{t}}(\hat{s},\hat{t},\hat{u})
\\ \nonumber
&&+f_{d/A}(x_{d},Q^2)f_{u/B}(x_{u},Q^2)
{d\hat{\sigma}\over d\hat{t}}(\hat{s},\hat{u},\hat{t})]
\end{eqnarray}
with
\begin{eqnarray}
\nonumber
x_{u}&=&\sqrt{\tau} e^{y^{b}} ,\qquad x_{d}=\sqrt{\tau} e^{-y^{b}} \\ \nonumber
\hat{s}&=&x_{u}x_{d}s, \qquad \hat{t}=-x_{u}p_{T}\sqrt{s}e^{-y_{1}},\qquad
 \hat{u}=-x_{d}p_{T}\sqrt{s}e^{y_{1}}\\ \nonumber
y^{\star}&=&(y_{1}-y_{2})/2 ,\qquad y^{b}=(y_{1}+y_{2})/2 \\ \nonumber
y_{1}^{\min}&=&\max (-Y, \log{{\tau}-y_{2}}) , \qquad y_{1}^{\max}=\min (Y, -\log{{\tau}-y_{2}}) , \qquad
\tau=M_{jj}^{2}/s \; . \\ \nonumber
\end{eqnarray}
In Eq. (9), the differential cross section for the subprocess 
$u d \to ud$ has the form 
\begin{eqnarray}
{d\hat{\sigma}\over d\hat{t}}(u d \to u d)&=&
{4\pi \over 9}[{\alpha_D^{2}(\hat{s}+\hat{t})^2\over
 \hat{s}^2[(\hat{s}-M_{D}^{2})^{2}+M_{D}^{2}\Gamma_D^{2}]} \nonumber \\
\nonumber \\
&&+{2\alpha_{s}^{2}\over
\hat{s}^{2}}{(2\hat{s}^{2}+\hat{t}^{2}+2\hat{s}\hat{t})\over\hat{t}^{2}}
-{2\alpha_D\alpha_{s}\over \hat{t}}{(\hat{s}+\hat{t})^2(\hat{s}-M_{D}^{2})
\over\hat{s}^2[
(\hat{s}-M_{D}^{2})^{2}+M_{D}^{2}\Gamma_D^{2}]} ]\qquad .
\end{eqnarray}
where the dominant interference with the  the t-channel gluon exchange 
is taken into account.

\begin{figure}[htb]
  \begin{center}
  \epsfig{file=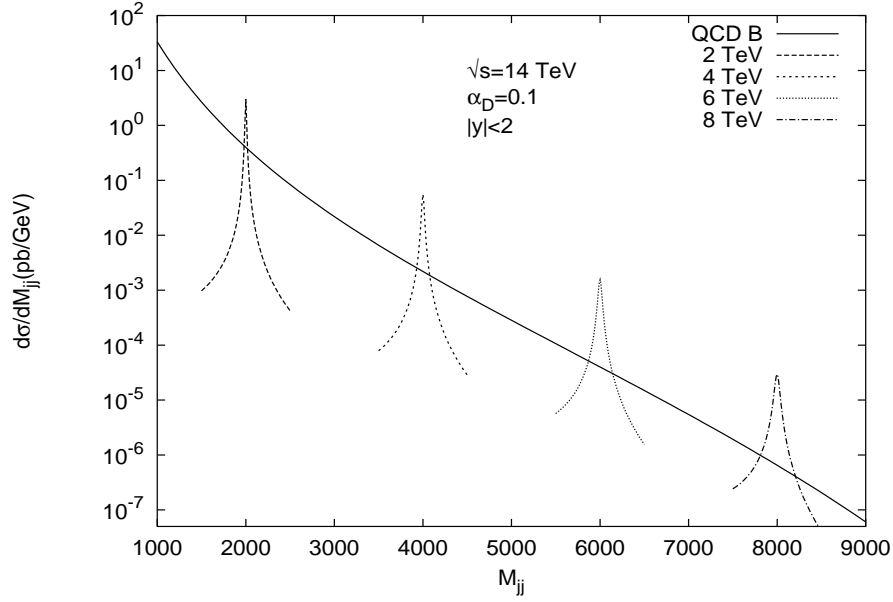,width=12cm,height=8cm}
  \end{center}
  \caption{Dijet invariant mass
distribution for  $M_D=2,4,6,8$ TeV superimposed over the QCD
background.} \label{fig2}
\end{figure}

Fig.~\ref{fig2} shows jet-jet invariant mass distribution for
the process $pp\to D_{2}\to jjX$ together with  the estimations of the
QCD backgrounds at LHC. For comparison, signal peaks for vector 
diquark masses $M_D=2,4,6,8$ TeV and
$\alpha_D=0.1$ are superimposed on the background distribution.

We have estimated background events for an integrated LHC
luminosity of $10^{5}$ pb$^{-1}$ taking into account, as an
example, the energy resolution of ATLAS hadronic calorimeter
\cite{atlas99}:

\begin{eqnarray}
{\delta E \over E}={50\%\over \sqrt{E}}+3\%
\end{eqnarray}
for jets with $\vert y\vert <3$. The mass resolution can be expressed 
as
\begin{eqnarray}
{\Delta M \over M}\approx {\delta E\over\sqrt{2}E} \qquad .
\end{eqnarray}
We have chosen $\Delta M$ as the mass window centered at $M_{D}$
for signal and background estimations. Here, we take the energy of
a jet $E\approx M_{D}/2$. For signal estimation the mass window
$\Delta M$ is taken to be 2$\Gamma_D$. This corresponds to the
95\% CL for statistical acceptance. The cross sections are
calculated by using the formula

\begin{equation}
  \sigma=\int_{M_D-\Delta M/2}^{M_D+\Delta M/2}
  dM_{jj}(\frac{d\sigma}{dM_{jj}})
\end{equation}

Number of signal ($S$) and background ($B$) events, and the
corresponding significances with $\alpha_D=0.1$ are tabulated in
Table ~\ref{table3}. Evidently, tighter cut on the rapidity $y$
improves the significance considerably. In Table ~\ref{table4}, we
present the achievable $M_D$ limits for different values of
$\alpha_D$. As discovery criteria, we adopt $S/\sqrt{B}\ge 5$ and
$S\ge 25$.

\begin{figure}[htb]
   \begin{center}
  \epsfig{file=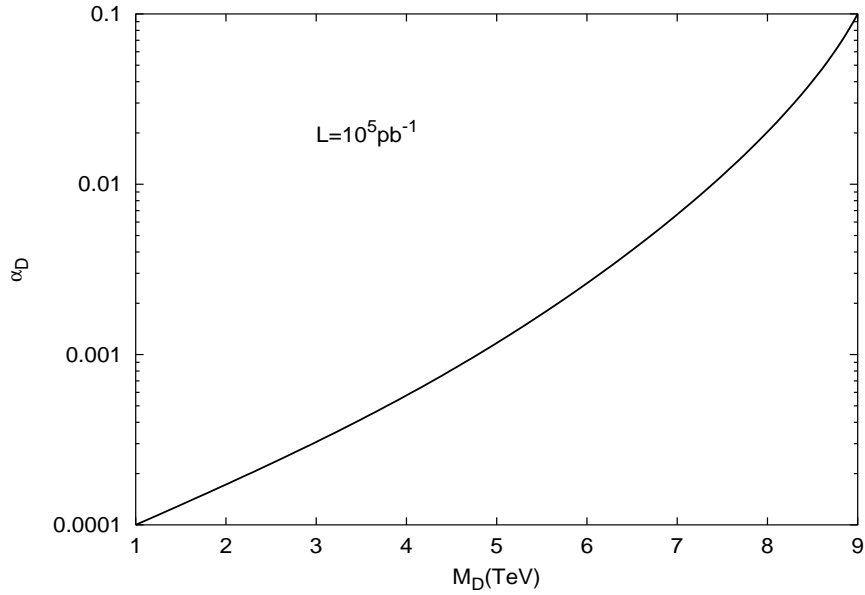,width=12cm,height=8cm}
  \end{center}
  \caption{Attainable limits for vector diquark $D_2$ in $\alpha_D$-$M_D$ plane.}
\label{fig3}
\end{figure}

\begin{figure}[htb]
  \begin{center}
  \epsfig{file=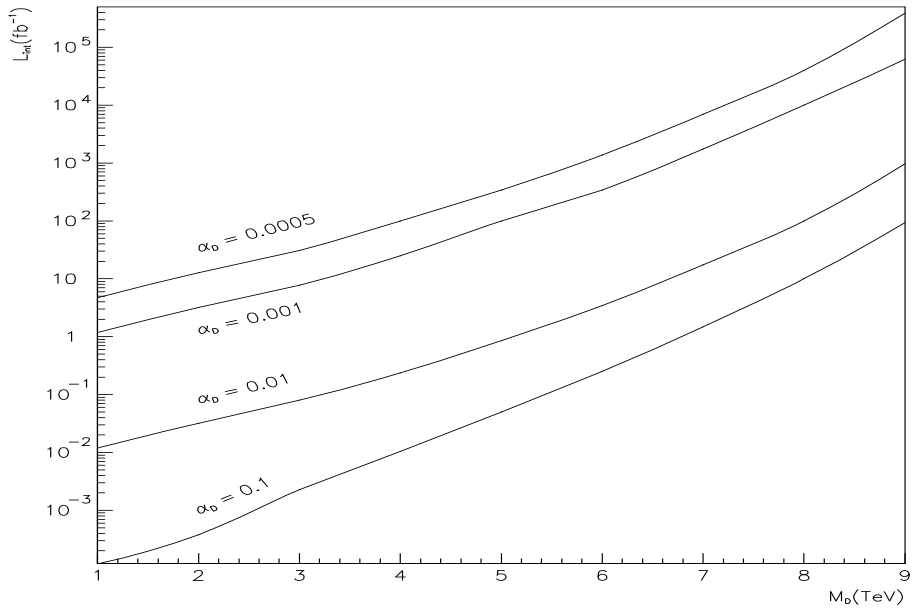,width=12cm,height=8cm}
  \end{center}
  \caption{The minimum integrated luminosities, needed to satisfy the adopted discovery
criteria, as a function of $M_D$ for various $\alpha_D$ values.}
\label{fig4}
\end{figure}

In Fig. \ref{fig3}, the attainable mass limits are presented in
the plane ($\alpha_D,M_D$). The LHC potential for the discovery of
$D_2$ is clearly demonstrated in Fig.~\ref{fig4} where minimum
integrated luminosities, needed to satisfy the adopted criteria,
are plotted as a function of $M_D$ for various $\alpha_D$ values.

There are ten different diquarks listed in Table 1, coupled to the first 
family quarks. All of them will decay into 2 jet final states. 
Vector and scalar type diquarks can be easily distinguished by the  
angular distribution of produced jets. 
To identify different vector (or scalar) diquarks at  hadron colliders, 
one needs polarized proton beams. 
In principle, electric charge of diquarks can be determined at future 
lepton and photon colliders, provided that the center of mass energy is 
larger than $2 M_D$. Furthermore, $\gamma p$ colliders based on linac-ring 
type $e p$ colliders will give essential contribution to the subject. 
The observation of the associated production of diquarks with leptoquarks will 
be possible at future lepton-hadron colliders.

In conclusion, the resonance production of vector diquarks at LHC has 
large cross
section. With reasonable cuts, it may be possible to cover mass
ranges up to $9$ TeV for coupling $\alpha_D=0.1$. For smaller
couplings as $\alpha_D=5\times 10^{-4}$, it is still possible to
probe diquarks up to the mass of $4$ TeV at an integrated
luminosity $L=10^2$fb$^{-1}$.

\newpage

\begin{table}[bth]
\caption{Quantum numbers of the first generation, color
$\bar{3}$ and baryon number $|B|=2 / 3$ diquarks
described by the effective Lagrangian (5) 
 according to $SU(2)_{W}\times U(1)_{Y}$ invariance.
 $Q_{em}=I_{3}+Y/2$
\label{table1}}
\begin{tabular}{ccccc}
\hline
Scalar Diquarks&SU(2)\( _{W} \)&U(1)\( _{Y} \)&\( Q_{em} \)&Couplings\\
\hline
\( D_{1} \)& 1 &2/3&1/3&\( u_{L}d_{L}(g_{1L}),\,u_{R}d_{R}(g_{1R}) \)\\
\hline 
\( \tilde{D}_{1} \)& 1& $-4/3$& $-2/3$&\( d_{R}d_{R}(\tilde{g}_{1R}) \)\\
\hline 
\( \tilde{D}^{\prime }_{1} \)& 1& 8/3&4/3&\( u_{R}u_{R}(\tilde{g}^{\prime }_{1R}) \)\\
\hline
\( D_{3} \)&3&2/3&\( \left( \begin{array}{c}4/3\\
1/3\\
-2/3
\end{array}\right)  \)&
\( \left( \begin{array}{c}
u_{L}u_{L}(\sqrt{2}g_{3L})\\
\begin{array}{c}
u_{L}d_{L}(-g_{3L})\\
d_{L}d_{L}(-\sqrt{2}g_{3L})
\end{array}
\end{array}\right)  \)\\
\hline
 Vector Diquarks&
&
&
&
\\
\hline
\( D_{2 } \)&2&$-1/3$&\( \left( \begin{array}{c}1/3\\
-2/3 
\end{array}\right)  \)&
\( \left( \begin{array}{c}
d_{R}u_{L}(g_{2})\\
d_{R}d_{L}(-g_{2})
\end{array}\right)  \)\\
\hline
\( \tilde{D}_{2 } \)&2&5/3&
\( \left( \begin{array}{c}
4/3\\
1/3
\end{array}\right)  \)&
\( \left( \begin{array}{c}
u_{R}u_{L}(\tilde{g}_{2})\\
u_{R}d_{L}(-\tilde{g}_{2})
\end{array}\right)  \)\\
\hline
\end{tabular}
\end{table}

\begin{table}[bth]
\caption{Cross sections (in pb) for QCD backgrounds contributing to 
2 jets final states
at parton level, generated by PYTHIA 6.1 with various $p_T$ cuts.
\label{table2}}
\begin{tabular}{lcccc}
\hline 
Process  & $p_T>100$ GeV & $p_T>500$ GeV & $p_T>1000$ GeV
&  $p_T>2000$ GeV  \\
\hline
$gg\to gg$ &6.3$\times 10^{5}$ &2.0$\times 10^{2}$
&2.3$\times 10^{0}$ &5.7$\times 10^{-3}$ \\
$q_{i}g\to q_{i}g $  &6.4$\times 10^{5}$
&4.8$\times 10^{2}$ &1.0$\times 10^{1}$
&5.7$\times 10^{-2}$\\
$q_{i}q_{j}\to q_{i}q_{j} $  &1.0$\times 10^{5}$
&1.8$\times 10^{2}$ &6.7$\times 10^{0}$ &8.8$\times 10^{-2}$ \\
$gg\to q_{k}\bar{q}_{k} $  &2.4$\times 10^{4}$ &9.8$\times 10^{0}$
&1.0$\times 10^{-1}$ &2.9$\times 10^{-4}$\\
$q_{i}\bar{q}_{i}\to q_{k}\bar{q}_{k} $  &1.6$\times 10^{3}$
&2.8$\times 10^{0}$ &1.3$\times 10^{-1}$ &1.1$\times
10^{-3}$ \\
$q_{i}\bar{q}_{i}\to gg $  &1.5$\times 10^{3}$ &2.5$\times 10^{0}$
 &6.7$\times 10^{-2}$  &8.5$\times 10^{-4}$\\
\hline
Total  &1.4$\times 10^{6}$ &8.8$\times 10^{2}$ &1.9$\times 10^{1}$ 
&1.5$\times 10^{-1}$ \\
\hline
\end{tabular}
\end{table}

\begin{table}[bth]
\caption{Observability of the vector diquark $D_2$
with $\alpha_D=0.1$ at LHC for $L_{int}=10^{5}$pb$^{-1}$.
\label{table3}}
\begin{tabular}{lccccc}
\hline
  $M_D$ (TeV) &1&3&5&7&9\\
\hline
$\vert y_{1,2}\vert < 2 $  \\
\hline
$S$ &7.7$\times 10^{7}$&1.1$\times 10^6$
&5.0$\times 10^4$ &1.7$\times 10^3$&2.7$\times 10^1$  \\
$B$ &2.8$\times 10^{8}$ &4.2$\times 10^5$
& 8.4$\times 10^3$ &2.2$\times 10^2$ &3.1$\times 10^0$ \\
$S/\sqrt{B}$ &4585 &1768 &545 &118 &16 \\
\hline
$\vert y_{1,2}\vert < 1 $ \\
\hline
$S$ &2.6$\times 10^7$ &5.8$\times 10^5$
&2.9$\times 10^4$ &1.1$\times 10^3$ &1.8$\times 10^1$ \\
$B$ &2.5$\times 10^7$ &4.0$\times 10^4$
&7.9$\times 10^2$ &2.1$\times 10^1$&3.0$\times 10^{-1}$ \\
$S/\sqrt{B}$ &5213 &2889&1044 &245 & 34 \\
\hline
\end{tabular}
\end{table}

\begin{table}[bth]
\caption{Achievable diquark mass limits for different quark-diquark-quark 
couplings in the framework of discovery criteria given in the text.
 $\vert y_{1,2} \vert<2$.
\label{table4}}
\begin{tabular}{ccccc}
\hline
$\alpha_D$&$M_D$(TeV) &S &B &S/$\sqrt{B}$ \\
\hline
0.1 &9  &27 &3.1 &16.0 \\
0.01 &8  &25 &30 &4.6 \\
0.001 &5  &500 &8400&5.4 \\
0.0005&4 &1200 &54000 &5.1 \\
\hline
\end{tabular}
\end{table}

\end{document}